# Wavelength Dependent Photocurrent of Hematite Photoanodes: Reassessing the Hole Collection Length

Asaf Kay, Daniel A Grave*, Kirtiman D Malviya, David S Ellis, Hen Dotan, and Avner Rothschild

Department of Materials Science and Engineering, Technion - Israel Institute of Technology, Haifa, Israel
*Email: dgrave@technion.ac.il

## Abstract

The photoelectrochemical behavior of a planar 1 μm thick Ti-doped hematite film deposited on F:SnO$_2$ coated glass was studied with both front and back illumination. Despite low quantum efficiency, photocurrent was observed upon back illumination with low wavelengths, indicating that some photogenerated holes are able to traverse at least 700 nm across the hematite film and effectively oxidize water. This cannot be accounted for using the commonly accepted hole collection length of hematite based on fitting to the Gartner model. Furthermore, under back illumination, 450 nm excitation resulted in increased photocurrent as compared to 530 nm excitation despite most of the light being absorbed further away from the surface. These results demonstrate that the photocurrent is strongly dependent on the optical excitation wavelength, and related to both delocalized holes with long lifetime and localized excitations rather than only being dependent on the proximity of the absorption to the surface.

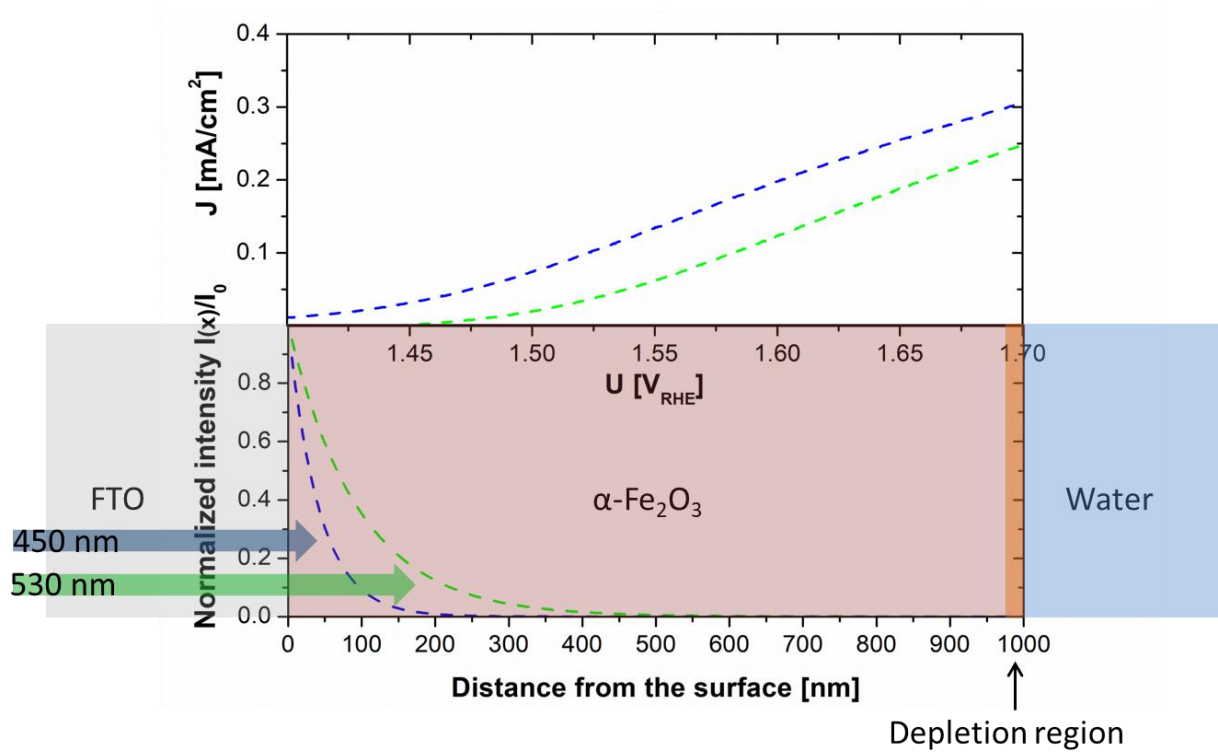

## Introduction

Hematite is a promising material for use as a photoanode in photoelectrochemical cells for solar water splitting due to its long term stability in alkaline solutions,[1] cost, abundance, and visible light absorption capabilities.[2] However, state of the art[3,4,5,6] photoanodes still fall significantly short of the theoretical efficiency. This poor performance is generally attributed to short lifetime[7] of photogenerated minority carriers, i.e. holes, primarily measured by optical pump-probe methods, and also to a reported small diffusion length,[8] resulting in significant bulk recombination. Despite much research into hematite photoanodes, an accurate model describing the device physics and charge transport of hematite photoanodes is still lacking. The most widely used model to describe photoanode behavior is the Gartner model,[9] which describes the photocurrent as a sum of all carriers photogenerated within the depletion region plus those generated in the bulk which are able to diffuse to the depletion region. It is widely accepted that hematite photoanodes display short depletion and diffusion lengths of only several nm,[8] therefore most of the efforts to improve hematite performance have centered around nanostructured porous layers.[5] In this work, we show that, in seeming contradiction to one or more of the preceding assumptions (i.e., Gartner model, diffusion length, or depletion region width), holes generated at least 700 nm away from the surface in a thick Ti-doped hematite planar photoanode are able to reach the surface and contribute to the photocurrent. Furthermore, we show that photogeneration of holes closer to the surface does not necessarily result in higher probability of charge carrier extraction and that the wavelength dependence of the generation plays a significant role in the ability to extract charges.

**Methods**

*Hematite Photoanode Fabrication.* Thick film (1 μm) of Ti doped hematite (α-$Fe_2O_3$) was deposited by pulsed laser deposition (PLD) on an FTO-coated glass substrate (Pilkington's TEC15). The film was deposited from a 1 cation% Ti doped hematite (α-$Fe_2O_3$) target prepared by solid state reaction route using high purity powders of $Fe_2O_3$ (99.99%, Alfa Aesar) and $TiO_2$ (99.995%, Alfa Aesar). The powders were mixed in appropriate amounts to obtain a Ti concentration of 1 cation%. The mixture was ball-milled for 24 h using YTZ milling balls (Tosoh, Japan), and subsequently pressed in a stainless steel mold and sintered in air at 1200°C for 12 h, resulting in a 1" disk-shaped pellet with a relative density of 88% from which the hematite film was deposited. Wavelength dispersive X-ray spectroscopy (WDS) analysis of the target confirms the nominal composition of 1 cation% Ti in $Fe_2O_3$.[10] The deposition was carried out using a PLD system equipped with a KrF (λ = 248 nm) excimer laser (COMPexPro 102, Coherent, GmbH). Before deposition the FTO substrate was rigorously cleaned to produce a high-quality film following the cleaning process reported elsewhere.[11] The film was deposited using 200,000 laser pulses with a fluence of 1 J/cm$^2$ and repetition rate of 15 Hz. The distance between the substrate and the target was 70 mm, and the set-point temperature was 500°C (at the heater) which corresponds to a substrate temperature of approximately 450°C. The deposition was carried out in $O_2$ atmosphere at a constant pressure of 25 mTorr.

*Materials characterizations.* X-ray diffraction (XRD) was used to identify the phase composition of the photoanode. The XRD diffractogram in Figure S1 was acquired using an X-ray diffractometer (Rigaku SmartLab) in parallel beam configuration with Cu Kα radiation in the range of 20-75° at a scan rate of 0.01°/s. Phase identification analysis confirms the presence of $Fe_2O_3$ hematite (α-$Fe_2O_3$, JCPDS 01-080-5413) and $SnO_2$ rutile phases (JCPDS 01-079-6887) arising from the film and substrate, respectively. The indexed diffraction pattern shows the film is strongly textured in (110) and (300) orientation.

High-resolution scanning electron microscopy (HRSEM, Zeiss Ultra Plus) was used to examine the surface morphology of the photoanode. Figure S2 presents a typical plan view HRSEM micrograph that shows a faceted growth of hematite film. Dual-beam focused ion beam (FIB, Strata 400S, FEI) was used to prepare a cross-section specimen for transmission electron microscopy (TEM), using the lift-out technique.[12,13] The film was coated with C and Pt for the lift out process. A monochromated and aberration-corrected TEM (FEI Titan 80-300 kV S/TEM) was used to examine the cross-section morphology of the hematite film on the substrate. Figure S3 presents a bright-field TEM micrograph of the cross-section specimen. Figure S3(a) shows the full stack of the hematite film deposited on a FTO-coated glass substrate with protecting C and Pt film during FIB sample preparation. The thick hematite film (1 µm) is dense, uniform and in a columnar structure. Figure S3(b) is the inset area of Figure S3(a) showing the sharp interface between the FTO and hematite layers.

Figure S4 presents a large area cross-section high-angle annular dark field (HAADF) micrograph of the specimen, obtained using the STEM mode. The selected area EDS elemental mapping of the region of interest (ROI) shows the compositional variation across the film stack. The composition map shows a sharp compositional contrast and no diffusion of Sn from FTO to the hematite film. The Ti doping concentration is low (1 cation%), not evident in the map. The separate EDS compositional analysis shows 1 cation% Ti doping in the hematite film (not shown here). Additional EDS measurements were carried out using high resolution SEM (Zeiss Ultra Plus) with an X-MAS detector (Oxford Instruments company). The measurement was carried out with a high voltage of 20 kV and a measurement time of 100 s. Five areas were sampled randomly to yield a 1.07 ± 0.06 cation% Ti doping.

Transmittance (T) and reflectance (R) spectra of the photoanode stack (glass/FTO/hematite) were measured using a spectrophotometer (Agilent Cary 5000), and the absorptance (A) was calculated using $A = 1 - T - R$. The spectra are shown in Figure S5.

*Photoelectrochemical measurements.* The photoelectrochemical performance of the photoanode was examined by three electrode voltammetry measurements with the photoanode serving as the working electrode and a platinum wire as the counter electrode. An Hg/HgO electrode in 1M NaOH solution served as the reference electrode. The photoelectrochemical measurements were carried out in alkaline aqueous electrolyte solutions (1 M NaOH in deionized water) without sacrificial reagents. Within some specific measurements a hole scavenger (0.5 M $H_2O_2$) was added to the 1 M NaOH solution (Figure S6). Current versus potential (*J-U*) linear sweep voltammograms were acquired both in the dark and in front light (from the film side) and back light illumination (from the glass side) using a solar simulator (ABET Technologies Sun 3000 class AAA solar simulator). In addition to the voltammetry measurements under solar simulated illumination, we also carried out measurements with 450 and 530 nm light emitting diodes (LED) using a Zahner Zennium electrochemical workstation (Figures S7-S9).

The incident photon conversion efficiency (IPCE) was measured using a 1000 W Newport Xenon lamp coupled to a monochromator as the light source, and Zahner potentiometer for applying potential and measuring current. The incident light intensity spectrum was measured using a power meter behind an aperture of identical size as that of the sample. Light intensity measurements were carried out before and after the voltammetry measurements to ensure stability of the incident intensity. Measurements were conducted under light bias provided by a white LED (Mightex "glacier white" 6500K) at 1000 mA current, incident at an oblique angle, on the same side of the sample as the monochromator light. In order to prevent errors due to drift which is especially prevalent under white bias, the current was measured for alternating conditions of shutter-on and shutter-off for several seconds each for each wavelength. An example is shown in Figure S10, measured at a high wavelength of 612 nm where the IPCE is expected to be small and yet, a clear photocurrent is observed. The IPCE measurements were also done without light bias (not shown), with almost an identical result for front illumination. For back illumination, a ~20-30% decrease was observed when not using light bias. The main error in the IPCE measurement is from the centering of the sample in the path of the monochromatic light which we estimate could scale the curve by ±10% of the reported IPCE. However, the ratio between the back and front values of the integrated IPCE - solar spectrum product measured at 1.65 V vs RHE is within 2% of the corresponding ratio between the back and front photocurrents measured at the same potential under solar simulated illumination. The absorbed photon conversion efficiency (APCE) was calculated by dividing the IPCE by the absorptance spectrum shown in Figure S5c. The APCE is plotted in Figure S11.

**Results and Discussion**

Figure 1 depicts a 1 μm thick hematite film deposited on an F:SnO$_2$ (FTO) current collector in front contact with an alkaline aqueous electrolyte. The region near the surface is termed the depletion layer or space charge region as it is depleted of majority charge carriers (electrons). Given that the FTO/Ti-doped hematite contact is reported to be Ohmic,[14,15] only a depletion region at the surface is considered. The depletion layer width is a function of the applied potential and dopant distribution across the photoanode. For heavily doped n-type hematite photoanodes working at high oxidative (anodic) potentials, the depletion layer width is typically less than 5-10 nm.[16] According to the Gartner model,[9] photogenerated charge carriers in this region are separated by the built-in electric field and collected to yield measurable photocurrent. Outside of this region, only minority charge carriers that are able to reach the depletion region through diffusion contribute to the photocurrent. The probability that a minority charge carrier in this region is able to reach the depletion region and be collected is given by $e^{(-x/L)}$, where x is the position of photogeneration relative to the depletion width and L is the minority charge carrier (hole) diffusion length. All the minority charge carriers generated in the bulk which are not able to reach the depletion region recombine and do not contribute to the photocurrent. The Gartner model has been modified to also account for recombination in the space charge region in the case of low mobility materials such as hematite.[17] The commonly cited diffusion length estimate of 2-4 nm for hematite was extracted by fitting photoelectrochemical voltammetry measurements on the basis of the Gartner model.[8] However, there has been no other independent verification of the hole diffusion length using direct measurement techniques.

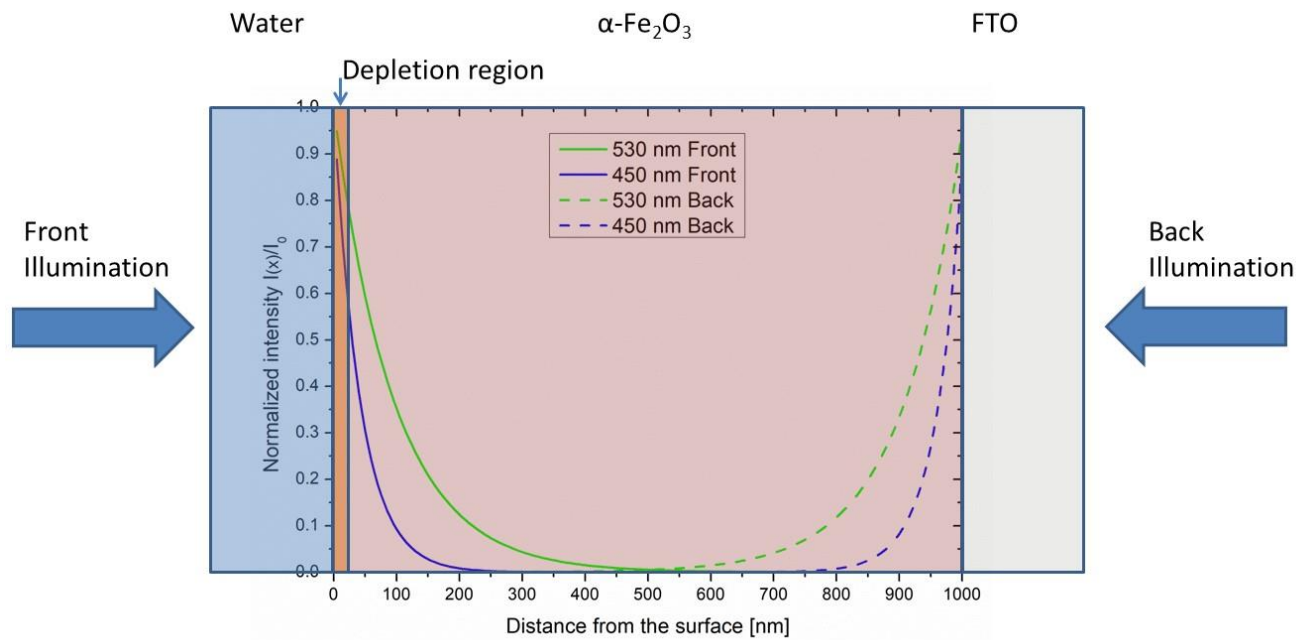

**Figure 1.** Schematic illustration of a 1 μm thick hematite photoanode with overlay curves showing light intensity profiles for both front and back illumination at wavelengths of 450 and 530 nm. The depletion region at the surface is not to scale (in reality it is only several nm thick).

A ~1 μm thick Ti-doped hematite film was deposited on a FTO-coated glass substrate by pulsed laser deposition (PLD). Fabrication details and material characterizations including X-ray diffraction (XRD), scanning electron microscopy (SEM), and transmission electron microscopy (TEM) can be found in the supplementary information. The theta-two-theta diffraction pattern shown in Figure S1 confirms that the film is phase pure hematite with a strong (110) and (300) preferred orientation. The SEM micrograph in Figure S2 shows a faceted and dense film surface without microcracks or porosity. This is further confirmed by the TEM cross section shown in Figure S3 that displays the dense columnar growth of the films. The FTO/hematite interface is sharp without evidence of diffusion, also confirmed by STEM and EDS analysis in Figure S4. While nanostructuring has been shown to improve photoelectrochemical performance,[5] the use of a compact film provides the ability to accurately model and define physical parameters such as the illumination profile and its relation to the photoanode/electrolyte junction in a one dimensional approach. This is especially critical when analyzing results in the context of models such as the Gartner model where the electrolyte should only be in contact with the front surface of the photoanode.

Figure 2 displays the linear sweep voltammogram of the ~1 μm thick Ti-doped hematite film in 1M NaOH using AAA solar simulated light under both front and back illumination. At high potentials, where surface recombination is minimized, the photocurrent observed from front illumination is roughly double that observed under back illumination. Such a high back illuminated photocurrent is a surprising result given the penetration depth of light into hematite and the large thickness of the films. To analyze the results in the context of the Gartner model, the depletion width and illumination profile must first be calculated. The depletion layer width, $W$, was calculated using the following formula, $W = ((2\kappa\epsilon_0 V_{bi})/(qN_d))^{1/2}$ where $\kappa$ is hematite dielectric constant of 33,[18] $\epsilon_0$ is the permittivity of free space, $V_{bi}$ is the built-in voltage calculated by subtracting the flat-band potential from the applied potential, q is the elementary charge, and $N_d$ is the donor density. The depletion width at a potential of 1.65 V vs. RHE, using the dopant density and flat-band potential extracted from Mott-Schottky measurements previously reported,[19] ranges from 1 - 7 nm (using values between 20 – 80 for the dielectric constant) for this film. Taking the reported diffusion length of 2-4 nm and adding it to the depletion layer thickness, we should expect collection length of approximately 10 nm. In order to calculate the amount of photons being absorbed in the front 10 nm of the film close to the front surface when the photoanode is illuminated from the back, the following formula was used: $\# photons = \int_{300\,nm}^{590\,nm} d\lambda \int_{990\,nm}^{1000\,nm} dx [I_{0(\lambda)}(1-R_{(\lambda)})\alpha_{(\lambda)}e^{-\alpha_{(\lambda)}x}]$ where R is the reflectance (measured by spectrophotometry), α is the extinction coefficient and $I_0$ is the spectrum of the AAA solar simulator used in the measurement. The generation in the front 10 nm amounts to $1.29\times10^{13}$ photons/cm$^2$. Even if all the photons generated in these 10 nm were converted to electron-hole pairs that were successfully separated and extracted, they would only contribute roughly 2 µA/cm$^2$ of photogenerated current, significantly lower than the measured ~400 µA/cm$^2$ under back illumination. Moreover, the back 700 nm of the film thickness would account for roughly 99% of the optical generation in the full 1 µm thick film. This suggests that the productive photogeneration of electron-hole pairs that are effectively separated and collected to oxidize water occurs at depths far from the surface, much farther than the estimated collection length.

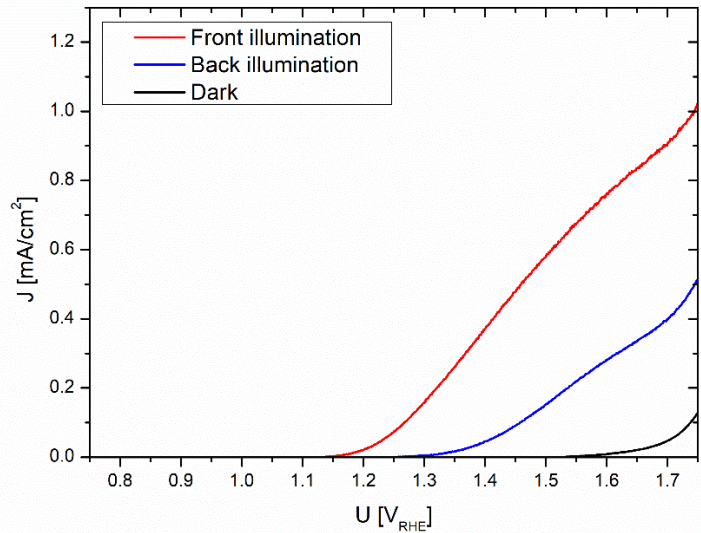

**Figure 2.** Front and back illuminated linear sweep voltammograms of 1 µm thick Ti-doped film in 1 M NaOH under solar simulated light.

To further investigate this phenomenon, we performed voltammetry measurements of the hematite photoanode in 1 M NaOH using light emitting diodes (LEDs) with emission wavelengths of 450 and 530 nm as the illumination sources. The calculated light intensity profiles at these wavelengths within the hematite photoanode are shown overlayed on the photoanode cross-section schematic in Figure 1. It can be observed that nearly all the photons at 450 and 530 nm are absorbed within a depth of ~300 and ~500 nm from the illuminated hematite interface, respectively. Figure S5 shows that the transmittance through the film at these wavelengths is zero as measured by UV-Vis spectrophotometry. Figure 3 shows the measured photocurrent for front and back illumination at 450 and 530 nm under a photon flux of 9.0x10$^{16}$ and 1.02x10$^{17}$ photons/s/cm$^2$, respectively. In front illumination, the photocurrent is higher for the shorter wavelength LED at similar photon flux. Assuming the same carrier generation per absorbed photon, a higher photocurrent for lower wavelengths would be expected based on the Gartner model since more of the photons absorbed under 450 nm illumination would generate carriers closer to the surface.

Under back illumination, non-negligible photocurrent is observed for both 450 and 530 nm illumination despite nearly all the light being absorbed in the first 300 or 500 nm, respectively, a surprising result. This data shows that holes generated as far as 700 nm from the surface can effectively oxidize water. Moreover, despite the fact that the 450 nm light is absorbed closer to the back interface and the photon flux is ~10% lower, it results in increased photocurrent as compared to the 530 nm light which is absorbed closer to the surface. This suggests that the probability of charge carrier separation and extraction depends on additional factors besides the proximity of the photogeneration to the surface. Furthermore, these results cannot be explained using both the Gartner model and a hole collection length of only several nm.

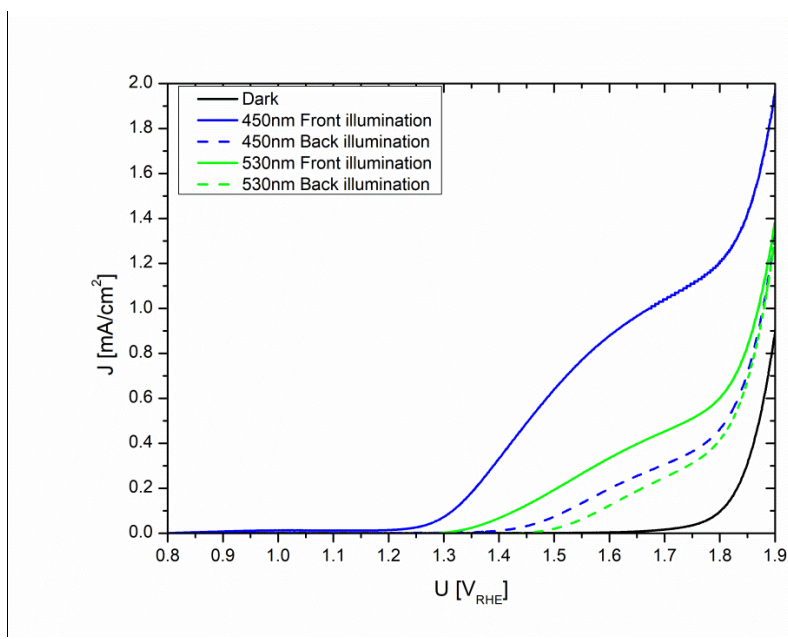

**Figure 3.** Front and back illuminated linear sweep voltammograms of 1 μm thick Ti-doped film in 1 M NaOH under LED emission wavelengths of 450 nm with photon flux of 9.0x10$^{16}$ and 530 nm with photon flux of 1.03x10$^{17}$ photons/s/cm$^2$.

To show that these results are indeed the cause of a bulk effect and not a surface effect, we performed measurements in hole scavenger (1M NaOH + 0.5M H$_2$O$_2$) solution where there is negligible surface recombination at all measured potentials.[20] These results are shown in Figures S6a and S6b for both solar simulated light and the different LEDs, respectively. The same qualitative behavior is observed in both hole scavenger solution and the 1 M NaOH pristine solution, confirming that the higher photocurrent for 450 nm illumination under back illumination is indeed related to a bulk effect. Additional linear sweep voltammograms for various photon fluxes were performed in both pristine and hole scavenger solutions (Figures S7-S8). These results show that the bulk photocurrent scales linearly with light intensity (Fig S9) at both wavelengths as expected for typical hematite photoanodes.[21]

The increased photocurrent for 450 nm illumination as compared to 530 nm can be explained on the basis of higher quantum efficiency at the lower wavelength. In order to study the wavelength dependent photoelectrochemical properties of the hematite film spectroscopically, incident photon conversion efficiency (IPCE) measurements were performed under both back and front illumination and are shown in Figure 4. The absorbed photon conversion efficiency (APCE), calculated by dividing the IPCE spectrum by the absorptance spectra, is plotted in Figure S11. Due to the large thickness of the film, the IPCE and APCE spectra are very similar.

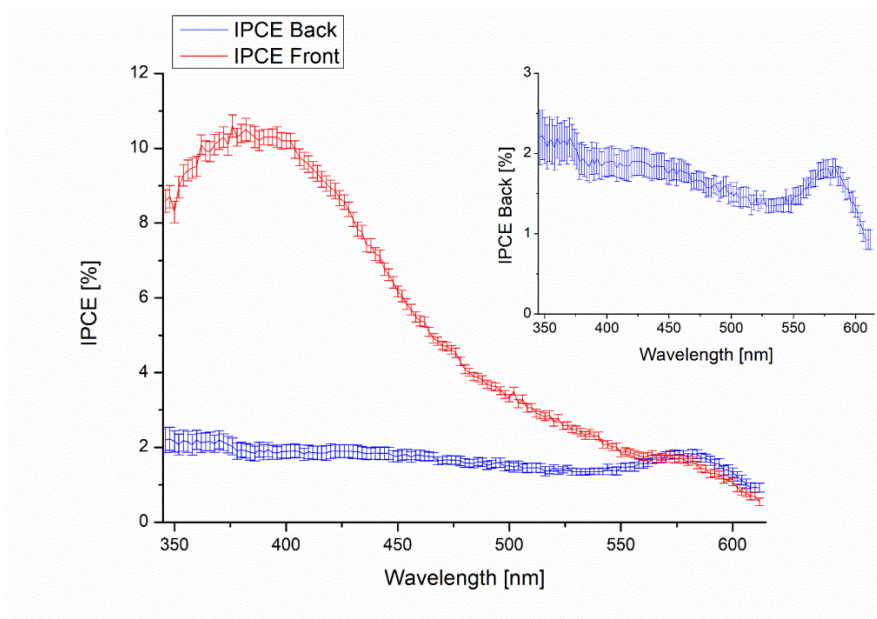

**Figure 4.** IPCE spectra of 1 µm thick Ti-doped film measured under both back and front illumination (blue and red curves, respectively) in 1 M NaOH at a potential of 1.65 V vs RHE. Inset shows the back illuminated IPCE spectrum on smaller y scale.

Under front illumination, the IPCE reaches a maximum of approximately 10% around 400 nm and then steadily decreases until 550 nm where another small peak emerges before the absorption edge at 600 nm. For back illumination, the IPCE is highest (2-3%) for the shortest wavelength of ~350 nm. Measurable IPCE at this wavelength suggests that photogenerated holes are able to traverse nearly the whole length (~1 µm) of the film. This is consistent with the LED measurements (see Methods). The IPCE decreases with increasing wavelength until around ~560 nm where an increase is observed, forming a peak around ~580 nm. This peak is not surprising given that light at high wavelengths (560 - 600 nm) can reach the surface in a 1 µm thick film as shown by the transmittance spectra in Figure S5 and the light intensity profiles calculated in Figure S12. The applied field can then assist in extraction of holes close to the surface. It has been shown that anodic bias can retard ultrafast recombination in hematite from timescales of few ps to longer timescales, leading to the presence of long lived holes on the hematite surface which can be extracted to oxidize water.[22,23]

If all of the absorbed photons were assumed to contribute equally to the photocurrent, then an increase in IPCE with increasing wavelength would be expected under back illumination across the whole spectrum, as more photons are being absorbed closer to the front surface. Since the opposite trend is observed, the spatial proximity of the absorption to the front surface alone cannot explain the observed results. Therefore, the wavelength dependence of the photogeneration of holes must also be a factor, responsible for the observed trends and peak like structures observed in the IPCE spectra. In a recent spectroscopic study, such peak structures were predicted in the context of specific absorption bands.[24]

The decrease in IPCE from 400 nm towards the band edge has been well documented for hematite photoanodes, although mostly for nanostructured ones,[25] and has generally been attributed to indirect *d-d* transitions which dominate the optical absorption spectra at high wavelengths and result in localized excitations.[26] It has been suggested that only two ligand to metal charge transfer (LMCT) bands are responsible for water-oxidation in hematite photoanodes and that the *d-d* bands which are responsible for most of the higher wavelength absorption are effectively inactive.[8,24] This may explain the peak structures observed in the IPCE spectra in the present report, and indeed the peak at 400 nm is consistent with ref [24]. A recent report has identified through 4D electron microscopy that illumination of hematite with a 519 nm laser generates $Fe^{4+}$ cations that have a lifetime of only a few picoseconds before returning back to the ground state ($Fe^{3+}$).[27] The authors suggest the existence of two types of holes, short lived $Fe^{4+}$ holes and longer-lived $O^-$ holes which are responsible for water photo-oxidation, consistent with other reports using transient absorption spectroscopy which have found that under 400 nm illumination, recombination processes can take several hundred picoseconds.[28] Another recent study of hematite used a model whereby small polaron trapping is responsible for carrier localization and that the polaron hopping radius and lifetime is strongly dependent on excitation wavelength with carriers being more localized at high wavelengths.[29] These studies show that the excited state dynamics in hematite is a complex and possibly multi-step process, involving multiple types of excitations and polaronic behavior. This continues to be the focus of ongoing theoretical work.[30] The wavelength dependent photocurrent observed in this work is consistent with the existence of both localized excitations and delocalized holes, as suggested in other studies.[8,24,27,29] Due to the complicated dynamics and various hematite preparation methods, there has been no consensus on the charge carrier lifetime, with processes observed in optical transient absorption spectra across a wide range of timescales and attributed to various phenomena.[7,22,23,24,28,29,31] Our results suggest that at least some of the delocalized holes can traverse distances significantly longer than previously thought.

Despite the ability of some photo-generated holes to traverse large distances, the low IPCE values measured in this work suggest that significant bulk recombination still prevails in these films. This is unsurprising given their polycrystalline nature and the high majority carrier concentration which are both expected to result in fast recombination. Possible routes of reducing recombination include the use of heterogeneous doping profiles to assist in charge separation[19] and growth of heteroepitaxial films with less crystalline defects.[32] Another route for improvement would be to increase the proportion of mobile charge carriers to localized excitations. For this, a better understanding of what governs the wavelength dependence and charge carrier dynamics in hematite is necessary for design of future photoanodes.

**Conclusions**

In conclusion, this work has directly shown that the collection length of holes in a 1 μm thick planar and dense (non-porous) Ti-doped hematite film photoanode is not limited to several nm as previously thought. Back illumination of the film with 450 and 530 nm excitations leads to measurable photocurrent despite all the light being absorbed within 300 and 500 nm of the back interface, respectively. This shows that photogenerated charges are collected from depths of at least 700 nm. Furthermore, the lower wavelength excitation results in increased photocurrent despite the absorption being further away from the front surface. This clearly highlights the importance of the wavelength dependence on the charge carrier generation, which, combined with the spatial dependence of the optical absorption, is a crucial factor for the collection efficiency. Our observations challenge previous assumptions about diffusion length in hematite, and open up new possibilities to improve the performance of hematite photoanodes. In parallel, effort should be focused to optimize the wavelength dependence of charge carrier generation in order to increase the proportion of mobile charge carriers to localized excitations.

**Supporting Information**

XRD of hematite photoanode; SEM, TEM, STEM and EDS characterization; optical spectra measured by UV-Vis spectrophotometer; photoelectrochemical hole scavenger measurements under 1 sun illumination and under 450 and 530 nm LEDs; photoelectrochemical measurements for different light intensities of 450 and 530 nm LEDs in both pristine 1 M NaOH and hole scavenger solutions; photocurrent as a function of photon flux for 450 and 530 nm light; IPCE measurements; and calculated APCE spectra.

**Acknowledgements**
This research has received funding from the European Research Council under the European Union's Seventh Framework programme (FP/200702013) / ERC Grant agreement n. [617516]. D. A. Grave acknowledges support by Marie-Sklodowska-Curie Individual Fellowship no. 659491. The results were obtained using central facilities at the Technion's Hydrogen Technologies Research Laboratory (HTRL), supported by the Adelis Foundation, the Nancy & Stephen Grand Technion Energy Program (GTEP) and by the Solar Fuels I-CORE program of the Planning and Budgeting Committee and the Israel Science Foundation (Grant n. 152/11).